# Infrared perfect absorber based on nanowire metamaterial cavities

Yingran He[1,2], Huixu Deng[1], Xiangyang Jiao[1], Sailing He[2], Jie Gao[1], and Xiaodong Yang[1,*]

[1] Department of Mechanical and Aerospace Engineering, Missouri University of Science and Technology, Rolla, MO 65409, USA

[2] Centre for Optical and Electromagnetic Research, Zhejiang Provincial Key Laboratory for Sensing Technologies, Zhejiang University, Hangzhou 310058, China

*Corresponding author: yangxia@mst.edu

An infrared perfect absorber based on gold nanowire metamaterial cavities array on a gold ground plane is designed. The metamaterial made of gold nanowires embedded in alumina host exhibits an effective permittivity with strong anisotropy, which supports cavity resonant modes of both electric dipole and magnetic dipole. The impedance of the cavity modes matches the incident plane wave in free space, leading to nearly perfect light absorption. The incident optical energy is efficiently converted into heat so that the local temperature of the absorber will increase. Simulation results show that the designed metamaterial absorber is polarization-insensitive and nearly omnidirectional for the incident angle.

Metamaterials with artificial structured composites can exhibit intriguing electromagnetic phenomena, such as negative refraction [1], invisible cloaking [2], near-zero permittivity [3], and indefinite cavities [4, 5]. The macroscopic properties of metamaterials can be tailored flexibly by designing the artificial meta-atoms. Typically, it is desirable to reduce the energy dissipation in metamaterials in order to enhance their optical performance. However, the energy loss in a metamaterial can also be utilized to achieve perfect light absorption. After the demonstration of the first metamaterial absorber in microwave frequency [6], absorbers working at higher frequencies such as THz and infrared range have been designed and demonstrated [7-11], which hold great promise in light harvesting, thermal detection and electromagnetic energy conversion.

Here a novel design of metamaterial absorber based on nanowire metamaterial cavities is proposed, as shown in Fig. 1. Periodic nanowire metamaterial cavities with a period of $P$ = 600 nm are grounded by a 150 nm thick gold film on a glass substrate. In each cavity, 6 by 6 gold nanowires with radius $r_0$ and center-to-center distance $D$ = 60 nm are embedded in alumina host with a cube geometry of $L$ = 360 nm and $h$ = 130 nm. Finite-element method (FEM) simulation is performed to obtain the reflection spectrum $R$. And the absorption spectrum $A$ is $A = 1 - R$, since the transmission is exactly zero due to the thick gold ground plane. The permittivity of gold is described by Drude model $\varepsilon_m(\omega) = 1 - \omega_p^2/\omega(\omega+i\gamma_0)$ with plasma frequency $\omega_p = 1.37 \times 10^{16}$ rad/s and bulk collision frequency $\gamma_0 = 4.08 \times 10^{13}$ rad/s. The permittivity of alumina is $\varepsilon_d = 3.0625$.

The absorption spectrum of nanowire metamaterial cavities array with $r_0$ = 12.5 nm at normal incidence is shown as the blue curve in Fig. 2(a). Perfect light absorption is achieved at 195.9 THz, together with a smaller absorption peak at 143.5 THz. According to the effective medium theory (EMT), the gold nanowires embedded in alumina host can be regarded as an anisotropic medium with the following permittivity components [12],

$$\varepsilon_x = \varepsilon_y = \varepsilon_d \frac{(1+f_m)\varepsilon_m + (1-f_m)\varepsilon_d}{(1-f_m)\varepsilon_m + (1+f_m)\varepsilon_d}, \quad \varepsilon_z = f_m\varepsilon_m + (1-f_m)\varepsilon_d \quad (1)$$

where $f_m = \pi r_0^2/D^2$ is the volume filling ratio of gold nanowire. The EMT simulated absorption spectrum is shown as the red curve in Fig. 2(a), and the slight shift of absorption peaks compared to the realistic nanowire structures is attribute to the weak nonlocal effects [13].

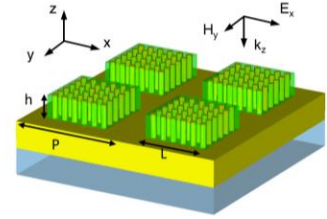

Fig. 1. Schematic of the nanowire metamaterial cavities array with a gold ground plane on a glass substrate. Each cavity is constructed with 6 by 6 gold nanowires (yellow color) embedded in alumina host (green color).

The magnetic field profiles $H_y$ in the $x$-$z$ plane of the cavity resonant modes are shown in Fig. 2(b), where the $m$ = 1 mode with one magnetic field peak along the $x$ direction is located at the perfect absorption frequency; while the $m$ = 3 mode corresponds to the weak absorption resonance. In the $x$-$y$ plane, the magnetic field profiles are homogeneous along the $y$ direction for both modes. The $m$ = 2 mode is absent at the normal incidence due to the structural mirror symmetry (with respect to $y$-$z$ plane) of the metamaterial cavities. The mechanism of perfect light absorption for the $m$ = 1 mode can be understood as the cavity mode supports the electric dipole resonance and the magnetic dipole resonance simultaneously. As shown in Fig. 2(c), the divergence and convergence of electric field at the top left and top right corners of the metamaterial cavity imply the accumulation of polarized positive charges and negative charges (with $\nabla \cdot \vec{E} = \rho/\varepsilon_0$, where $\rho$ is the polarized charge density), which serves as a strong electric dipole. Fig. 2(d) shows the distribution of displacement current $D$, with $\partial D/\partial t = -i\omega D = -i\omega\varepsilon E$. Strong displacement current is found inside the grounded gold film due to the large negative permittivity of metal.

Anti-parallel displacement currents are formed between the nanowire cavity and the gold ground plane, resulting in a strong magnetic dipole resonance, which is also shown in Fig. 2(b). To give a quantitative description of the electric and magnetic resonance, effective parameter retrieval is performed to obtain the effective permittivity $\varepsilon_{eff}$ and the effective permeability $\mu_{eff}$ of the cavity mode. Since the information of transmission is necessary for parameter retrieval [14], a ground film thickness 30 nm is used to allow an ultra-small transmission while the reflection is almost unaffected. $\varepsilon_{eff}$ = 0.401+3.456i and $\mu_{eff}$ = 0.609+3.229i at the $m$ = 1 resonance frequency are achieved, which leads to a cavity mode impedance of $Z = \sqrt{\mu_{eff}/\varepsilon_{eff}}\, Z_0$ = ( 0.942 - 0.067i )$Z_0$, matching to the free space impedance $Z_0$. The huge imaginary parts of $\varepsilon_{eff}$ and $\mu_{eff}$ are due to the strong optical loss in the metal nanowires, resulting in the perfect light absorption. The current absorber works in a narrow bandwidth due to the resonance nature of light absorption. Broadband absorption can be designed with either mixed multiple resonators or single multimode resonator [10, 11].

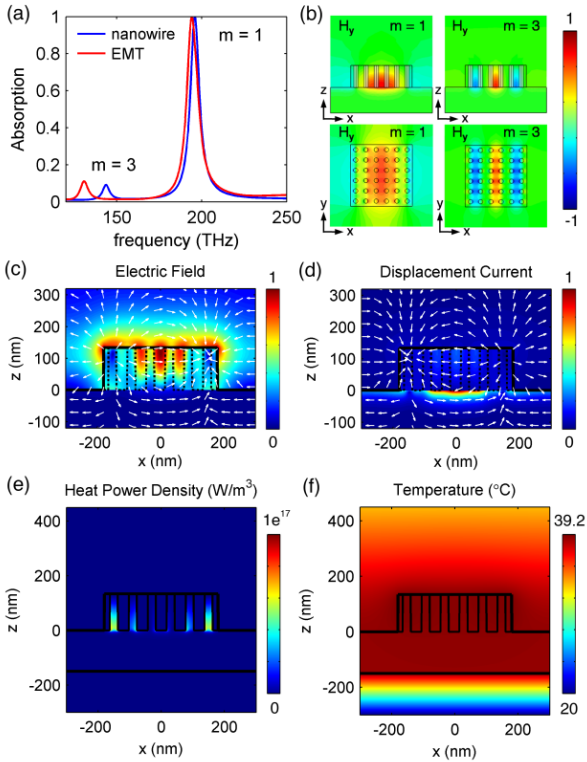

Fig. 2. (a) The absorption spectrum of nanowire metamaterial cavities array with $r_0$ = 12.5 nm at normal incidence. (b) The magnetic field profiles of the $m$ = 1 and $m$ = 3 resonant modes. The intensity and direction of (c) the electric field and (d) the displacement current for $m$ = 1 mode at $y$ = 0. (e) The heat power density and (f) the temperature distribution for $m$ = 1 mode at $y$ = 30 nm.

The absorbed light will be eventually converted into heat due to the resistive loss of gold nanostructure, which works as a nanoscale heat source to increase the local temperature inside the metamaterial cavities. As shown in Fig. 2(a), the absorption quality factor is about 20, which means the optical energy will be transferred to electrons within 20 optical cycles (~100 fs). Then electron-lattice relaxation occurs, giving rise to the heating of gold nanostructure. Subsequently, heat is conducted to surrounding materials, and eventually reaches the thermal equilibrium. At the thermal equilibrium, heat transfer equation $\nabla\cdot(-k\nabla T) = q$ can be solved to achieve the temperature distribution $T$, where $k$ is the thermal conductivity and $q$ is the heat power density generated from the light absorption, $q(\boldsymbol{r}) = (\omega/2)\mathrm{Im}[\varepsilon(\omega)]\varepsilon_0|\boldsymbol{E}(\boldsymbol{r})|^2$. With a background temperature of 20 °C, Fig. 2(e) and Fig. 2(f) show the distributions of the heat power density and the temperature, when the absorber is irradiated by incident light with an optical power intensity of 138.9 μW/μm² (corresponding to a incident power of 50 μW per unit cell area). Since heat generation is proportional to $\mathrm{Im}[\varepsilon(\omega)]|\boldsymbol{E}(\boldsymbol{r})|^2$, all the heats are generated inside the gold nanostructure, while the strongest heat power density is located at the bottom of the left and right nanowires where the electric field is largest, as shown in Fig. 2(e). The generated heat will be conducted away from gold nanowires into the surrounding alumina, air and glass substrate, leading to the increase of local temperature as high as 19.2 ℃, as plotted in Fig. 2(f). Since the thermal conductivity of gold is much larger than its surrounding materials, an almost uniform temperature distribution is observed inside the gold structures and a large temperature gradient is located in the glass substrate. The temperature variation is related to the incident optical power linearly.

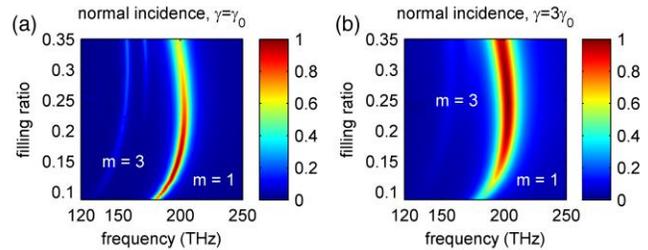

Fig. 3. The dependence of absorption spectra on metal filling ratio at normal incidence with different collision frequencies of gold, (a) $\gamma = \gamma_0$ and (b) $\gamma = 3\gamma_0$.

The impedance matching condition is critical for obtaining perfect light absorption, so it is expected that $f_m$ will largely affect the performance of the metamaterial absorber. Fig. 3 shows the dependence of absorption spectra on $f_m$ at normal incidence with different collision frequencies of gold $\gamma$, where the variation of filling ratio is derived by tuning the radius of the gold nanowires from $r_0$ = 10 nm to $r_0$ = 20 nm. Considering that the collision frequency $\gamma$ increases a lot due to the inevitable surface roughness of the fabricated sample, numerical simulations with $\gamma = \gamma_0$ and $\gamma = 3\gamma_0$ are shown in Fig. 3(a) and Fig. 3(b). For $\gamma = \gamma_0$, perfect absorption is obtained around $f_m$ = 0.14. For $\gamma = 3\gamma_0$, perfect absorption occurs at a larger filling ratio around $f_m$ = 0.25. The absorption spectra of $\gamma = 3\gamma_0$ show much broader bandwidth compared to that of $\gamma = \gamma_0$. The existence of optimized metal filling ratio for the perfect absorption can be understood as the following: when $f_m$ is smaller than the optimized value, the imaginary parts of the $\varepsilon_{eff}$ and $\mu_{eff}$ corresponding to the nanowire structure are not sufficiently large to totally absorb the incident light; when $f_m$ is larger, the impedance cannot match to the free space, both of which result in low light absorption. By comparing the absorption spectra of the two collision frequencies, it is also found that the resonance for the $m = 3$ mode becomes much weaker at $\gamma = 3\gamma_0$ since high order resonance is very sensitive to the material loss. Besides, there is an additional resonance peak between the $m = 1$ and the $m = 3$ modes at large filling ratios ($f_m > 0.22$) when $\gamma = \gamma_0$. This is a high order mode oscillating along the $y$ direction.

Since the metamaterial cavities array possesses 4-fold rotation symmetry in the $x$-$y$ plane, it is polarization-independent at normal incidence. For oblique incidence, however, the absorption performance will depends on the polarization. The incident electromagnetic field configurations of transverse electric (TE) polarization or transverse magnetic (TM) polarization are shown in Fig. 4(a) and Fig. 4(b). And the absorption spectra for both polarizations at two different collision frequencies are shown in Fig. 4(c-f). The absorption of TE polarized light will decrease gradually as the incident angle $\theta$ increases, while the peak absorption frequency remains the same. The absorption of TM polarized light, on the other hand, decreases in a slower way than that of TE polarized light, while the peak absorption frequency slightly shifts with the growth of incident angle. Nevertheless, the absorption remains strong for incident angle up to 80° for both polarizations. An additional absorption peak is noted for incident angle larger than 40° in Fig. 2(d) and 2(f), which is the $m = 2$ mode excited by the oblique incident light.

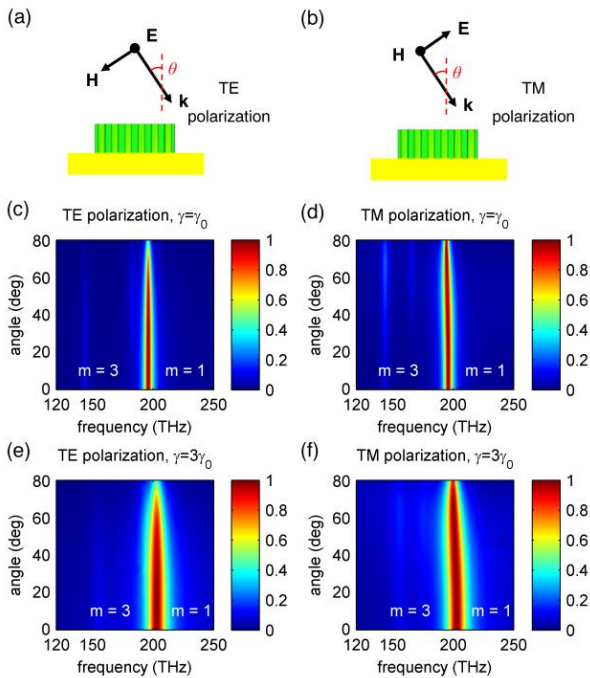

Fig. 4. (a), (b) The configurations of TE polarization and TM polarization. (c - f) The dependence of light absorption on the incident angle $\theta$ for different polarizations and collision frequencies with $f_m = 0.14$ in (c, d) and $f_m = 0.25$ in (e, f).

In conclusion, an infrared perfect absorber based on nanowire metamaterial cavities array is proposed and demonstrated. The cavity resonant mode with mode order $m = 1$ is utilized to excite both the electric dipole and the magnetic dipole simultaneously. At a specific frequency, the impedance of the metamaterial absorber can match to the free space, and the strong optical loss in nanowire structures results in the total absorption of the incident light, which generates heat and increases the local temperature in the metamaterial absorber. The influences of metal filling ratio and metal collision frequency on the absorption performance are also investigated. The metamaterial absorber can work in broad incident angles for both the TE and TM polarizations, which can be applied in the field of thermal absorption and emission, light harvesting, photodetecting, and electromagnetic energy conversion.

This work was partially supported by the Department of Mechanical and Aerospace Engineering, the Materials Research Center, the Intelligent Systems Center, and the Energy Research and Development Center at Missouri S&T, the University of Missouri Research Board, the Ralph E. Powe Junior Faculty Enhancement Award, and the National Natural Science Foundation of China (61178062 and 60990322).

References

1. R. A. Shelby, D. R. Smith, and S. Schultz, Science **292**, 77-79 (2001).
2. J. B. Pendry, D. Schurig, and D. R. Smith, Science **312**, 1780-1782 (2006).
3. A. Alù, M. G. Silveirinha, A. Salandrino, and N. Engheta, Phys. Rev. B **75**, 155410 (2007).
4. J. Yao, X. Yang, X. Yin, G. Bartal, and X. Zhang, Proc. Natl. Acad. Sci. **108**, 11327-11331 (2011).
5. X. Yang, J. Yao, J. Rho, X. Yin, and X. Zhang, Nat. Photon. **6**, 450-454 (2012).
6. N. I. Landy, S. Sajuyigbe, J. J. Mock, D. R. Smith, and W. J. Padilla, Phys. Rev. Lett. **100**, 207402-207404 (2008).
7. H. Tao, N. I. Landy, C. M. Bingham, X. Zhang, R. D. Averitt, and W. J. Padilla, Opt. Expr. **16**, 7181-7188 (2008).
8. N. Liu, M. Mesch, T. Weiss, M. Hentschel, and H. Giessen, Nano Lett. **10**, 2342-2348 (2010).
9. Y. Avitzour, Y. A. Urzhumov, and G. Shvets, Phys. Rev. B **79**, 045131 (2009).
10. X. Liu, T. Tyler, T. Starr, A. F. Starr, N. M. Jokerst, and W. J. Padilla, Phys. Rev. Lett. **107**, 045901 (2011).
11. Y. Cui, K. H. Fung, J. Xu, H. Ma, Y. Jin, S. He, and N. X. Fang, Nano Lett. **12**, 1443-1447 (2012).
12. A. Sihvola, *Electromagnetic mixing formulas and applications* (Institution of Electrical Engineers, 1999).
13. J. Elser, V. A. Podolskiy, I. Salakhutdinov, and I. Avrutsky, Appl. Phys. Lett. **90**, 191109 (2007).
14. X. Chen, T. M. Grzegorczyk, B.-I. Wu, J. Pacheco, Jr., and J. A. Kong, Phys. Rev. E **70**, 016608 (2004).